\documentclass[trackchanges,twocolumn]{aastex701}

\usepackage{graphicx}%
\usepackage{amsmath,amssymb,amsfonts}%
\usepackage{mathrsfs}%
\usepackage{xcolor}%
\usepackage{listings}%
\usepackage{tabularx}
\usepackage{multirow}
\usepackage{booktabs}
\usepackage{array}
\newcolumntype{Y}{>{\raggedright\arraybackslash}X}
\renewcommand{\arraystretch}{1.2} 

\begin{document}

\title{Unraveling the emission mechanism powering long period radio transients\\ from interacting white dwarf binaries via kinetic plasma simulations}

\author[orcid=0000-0003-0805-8234,gname=Yici,sname='Zhong']{Yici Zhong}
\affiliation{TAPIR, Mailcode 350-17, California Institute of Technology, Pasadena, CA 91125, US}
\affiliation{Walter Burke Institute for Theoretical Physics, California Institute of Technology, Pasadena, CA 91125, US}
\email[show]{yczhong@caltech.edu}  

\author[orcid=0000-0002-0491-1210,gname=Elias, sname='Most']{Elias R. Most} 
\affiliation{TAPIR, Mailcode 350-17, California Institute of Technology, Pasadena, CA 91125, US}
\affiliation{Walter Burke Institute for Theoretical Physics, California Institute of Technology, Pasadena, CA 91125, US}
\email{emost@caltech.edu}

\begin{abstract}

Recent observations of long period radio transients, such as GLEAM-X J0704–37 and ILTJ1101 + 5521, have revealed a previously unrecognized population of {galactic} radio transient sources associated with white dwarf--M dwarf binaries. 
{It is an open question how to produce coherent radio emission in these systems, though a model driven by binary interaction seems likely given the nature and correlation of the emission with the binaries' orbital period.}
{Using kinetic plasma simulations,} we demonstrate that the {relativistic} electron cyclotron maser instability (ECMI) is a viable mechanism for generating radio pulses in white dwarf–M dwarf systems, akin to planetary radio emission, such as that from the Jupiter-Io system.  
{We quantify the relativistic ECMI in the nonlinear regime under conditions relevant for white dwarf radio emission for the first time. Our simulations demonstrate that the ECMI can intrinsically produce partially linearly polarized emission relevant to explaining the observed emission spectrum of the two galactic sources, though the precise details will depend on the plasma composition.
Our work paves the way for a systematic and fully nonlinear computational modeling of radio emission from interacting white dwarf sources. }

\end{abstract}

\keywords{
\uat{Galactic radio sources}{571};
\uat{High Energy astrophysics}{739};
\uat{Interacting binary stars}{801};
\uat{M dwarf stars}{982};
\uat{Magnetospheric radio emissions}{998};
\uat{White dwarf stars}{1799};
\uat{Radio transient sources}{2008}
}

\section{Introduction}

Radio emitting white dwarfs (WD) pose fascinating challenges to our understanding of coherent radio emission mechanisms. Most famously the WD -- M-dwarf (MD) binary AR Scorpii has been shown to emit pulsed multi-wavelength (incl. radio) emission at a rate consistent with the rapid spin of its WD component  and extremely strong magnetic fields reaching up to $10^8\, \rm G$ \citep{2017NatAs...1E..29B}. Last year, another class of radio emitting galactic WD-MD sources was discovered, featuring (ultra-)long period radio transients (ULPTs).  These include ILT J1101+5521 at a distance of 340 pc \citep{deRuiter:2024yib} and GLEAM-X J0704-37 \citep{2024ApJ...976L..21H} at a distance of 380 pc \citep{Rodriguez:2025spg}. Both of these sources feature minute-long radio pulses with hours-long periods, and have been spectroscopically confirmed to host MDs \citep{deRuiter:2024yib,Rodriguez:2025spg}.
From an evolutionary point of cataclysmic variables (CVs), these close orbits imply a nearly synchronized state of the system \citep{2021NatAs...5..648S}, characteristic of (pre-)polars -- WDs with extreme magnetic fields \citep{2015SSRv..191..111F}.
While other LPTs with shorter periods have been discovered, as well, recently (e.g., \citealt{2025ApJ...988L..29D,Bloot:2025bgq}), it remains an open question whether all LPTs are WD hosting, or can also be powered by neutron stars (NS), especially for systems with X-ray counterparts \citep{2025Natur.642..583W}, or significantly higher luminosities than the confirmed WD-MD systems \citep{2025ApJ...988L..29D}.
What makes the ULPT sources above appealing from a theoretical point, is that their radio emission is very likely powered by orbital motion instead of spin down, making their radio emission more like those of planetary systems, akin to processes in our own Solar system (see also \citet{Horvath:2025sze} for alternative stellar wind powered scenarios).

Radio emission in planet-moon and star-planet systems is well understood \citep{Callingham:2024brj}. Observationally, Jupiter's decametric radio emission has been long studied \citep{warwick1964radio}, and shows clear modulations associated with Io's orbit \citep{1964Natur.203.1008B}. Theoretically, this process is well modeled with a unipolar inductor scenario \citep{1969ApJ...156...59G}, where the orbital motion of Io creates a closed DC current (see \citet{1993Sci...262.1035C} for observational confirmation), leading to coherent radio emission as outlined below. 
The details of particle acceleration in this system are likely complicated, and require mechanisms such as turbulence to convert large scale Alfven waves into kinetic ones \citep{2023RvMPP...7....6L} (see, e.g., \citet{2020GeoRL..4788432S} for recent JUNO observations of this process), followed by subsequent electron Landau damping \citep{2018JGRA..123.9560S}.
Similar mechanisms have also been proposed for orbiting WD and WD-planet systems \citep{Willes:2003mw,DallOsso:2005jkl}. This mechanism assumes that the exterior magnetic field of the primary planet efficiently threads the secondary companion. While this mechanism in most cases is insufficient to drain large parts of the orbital energy and alter the evolution of the system \citep{Lai:2012qe} (see also \citet{Piro:2012rq}), the close proximity of both planetary and galactic WD-MD sources compensates for this intrinsically lower luminosity. One crucial aspect of any such interaction is that it assumes that the relative twists incurred in the connecting flux tube are small. Otherwise, the connected flux tubes can rapidly expand and flare, disrupting the connection \citep{Lai:2012qe}. In the case of neutron star binaries, this has been demonstrated to lead to Fast Radio transients \citep{Most:2020ami,Most:2022ojl,Most:2023unc,Skiathas:2025pnj}, though the likely coherent radio emission mechanism for this specific process cannot operate in the WD context \citep{Most:2022ayk} (see also \citet{Lyubarsky:2020qbp,Mahlmann:2022nnz}).

One important aspect of ULPTs is the production of coherent radio emission. Planetary systems are subject to various types of radio emission, including Langmuir-wave driven type III radio bursts \citep{2014RAA....14..773R}, and cyclotron masers \citep{Chu:2004zz}. Especially the electron cyclotron maser instability (ECMI) is commonly invoked to explain companion-driven planetary radio emission \citep{1982ApJ...259..844M}, including in accreting WD systems \citep{1982ApJ...255L.107C,1984MNRAS.208..865S} (see also \citet{Qu:2024gng} for a recent discussion in the context of WD-MD systems). 
Extensive analytical work describing the conditions for linear growth, saturation and other properties of the ECMI has been carried out \citep{1958AuJPh..11..564T,1979ApJ...230..621W,1982ApJ...259..844M}, though few direct numerical simulations of the instability exist \citep{2011A&A...526A.161K,2012A&A...539A.141K,2011PhPl...18i2110L,2020ApJ...891L..25N,2021ApJ...909L...5L,2024A&A...681A.113L}, see also \citet{2023PhRvL.130p5101B,2025SciA...11.8912B} for simulations in the context of laser plasmas. Such simulations are important to fully assess and study the nonlinear regime of the instability.
Paralleling efforts to understand the potential role of (shock-powered) synchrotron maser emission in the context of Fast Radio Bursts \citep{Beloborodov:2019wex,Metzger:2019una} has highlighted the importance numerical kinetic simulations play in elucidating the radio energy conversion efficiency and polarization properties of the resulting emission \citep{Plotnikov:2019zuz,Babul:2020sil,Sironi:2021wca,Vanthieghem:2024van}. This is particularly important as the observed ULPTs feature strong linear polarization not commonly associated with the ECMI in analytical works, but recently demonstrated in a different context using kinetic simulations \citep{2025SciA...11.8912B}.

Overall, the continued discovery of ULPT sources with more complex observational constraints necessitates a full kinetic modeling to identify and confirm the likely emission mechanism in this system \citep{Qu:2024gng}.
Using fully kinetic particle-in-cell (PIC) simulations of collisionless plasma dynamics we provide the first numerical investigation of the ECMI in the mildly relativistic regime, which is appropriate for WD-MD systems. Our simulations reveal linear polarization properties in line with the observations, and provide a full quantification of the radio conversion efficiency, and spectral width of the instability.

\section{Basic picture}\label{sec:theory}

In this work, we want to provide a detailed account and explanation of ECM emission powered by interacting WD binaries, which has become especially urgent following the detection of several galactic radio transient sources confirmed as WD-MD systems \citep{deRuiter:2024yib,2024ApJ...976L..21H}. This scenario of radio emission driven by orbital motion of a binary system was first proposed by \citet{1969ApJ...156...59G}, and later extended by \citet{Willes:2003mw} and \citet{Lai:2012qe} for the case of interacting WD systems~\citep[see also][for a review of exoplanet scenarios]{Callingham:2024brj}.
We begin by providing a brief summary of the overall picture, before giving a detailed account of our kinetic simulation results.

Consistent with the observations of ILTJ1101 + 5521, we assume that the MD magnetic field is too weak to explain the observed radio luminosity, in addition to strong constraints on polarization typically observed in MD \citep{deRuiter:2024yib}. Consequently, we assume that only the WD has a magnetic field, which we take to be a dipole field, $B = B_\ast \left(R_{\rm WD}/R\right)^3$, where $B_\ast = \mu/R_{\rm WD}$ is the surface magnetic field of the WD, $R_{\rm WD}$ its radius and $\mu$ the WD magnetic moment. The binary is orbiting with an angular frequency $\Omega$, and the spin, $\Omega_{\rm WD}$, of the WD is such that the deviation from corotation $\Delta \Omega = \left|\Omega-\Omega_{\rm WD}\right| \ll 1$ is small. We denote the distance to the system by $d$.

In the ECM picture, the orbital motion of the WD-MD system drives a current on the flux tube connecting the WD and the MD. This is consistent with the observed period of the emission being up to two orders of magnitude larger than the spin down period of WD, e.g., as observed in AR Scorpii \citep{2017NatAs...1E..29B}.
The connected flux tube will form a closed circuit with dissipation, where the energy is provided by the orbital motion \citep{1969ApJ...156...59G}. 
We can compute a bound on energy dissipation rate, $\dot{E}_{\rm diss}$, using the unipolar inductor model \citep{Lai:2012qe},
\begin{align}
\dot{E}_{\rm diss} = & \left(2 \pi\right)^{13/3}\frac{\zeta_\phi}{2} \left(\frac{\Delta \Omega}{\Omega}\right) \frac{\mu^2 R_{\rm MD}^2} {\left[G (M_{\rm MD} +M_{\rm WD}) \right]^{13/3}}\,, \nonumber\\
=&\, 6.0 \times 10^{29} \mathrm{erg\ s^{-1}} \, \zeta_\phi 
\left( \frac{\Delta \Omega}{\Omega} \right)
\left( \frac{\mu}{10^{33}\, \mathrm{G\, cm}^3} \right)^2
\nonumber\\
&\left( \frac{R_c}{0.217 R_\odot} \right)^2 \left( \frac{M_\star + M_c}{0.8\, M_\odot} \right)^{-5/3}
\left( \frac{P}{100\, \mathrm{min}} \right)^{-13/3},
\label{eq:diss}
\end{align}
here $\zeta_\phi$ encapsulates the resistive efficiency of dissipation in the closed circuit. Importantly, $\zeta_\phi <1$, since otherwise net twists can built up in the connected flux tube leading it to expand and flare \citep{Most:2020ami,Most:2022ojl,Most:2023unc}. 
This dissipation provides an upper limit to the radio luminosity, $\mathcal{L}_{\rm radio}$, by assuming a radio emission conversion efficiency, $\xi$, such that 
\begin{align}
\mathcal{L}_{\rm radio} &\simeq \xi \dot{E}_{\rm diss} \,, \nonumber\\
&\simeq 6.0 \times 10^{27}  \mathrm{erg\ s^{-1}} \, \zeta_\phi 
\left( \frac{\Delta \Omega}{\Omega} \right) \left( \frac{P}{100\, \mathrm{min}} \right)^{-13/3} \nonumber
\\
& \left(\frac{\xi}{10^{-2}}\right) \left( \frac{\mu}{10^{34}\, \mathrm{G\, cm}^3} \right)^2 \left( \frac{M_\star + M_c}{0.8\, M_\odot} \right)^{-5/3}\nonumber \\
&\left( \frac{R_{\rm MD}}{0.217 R_\odot} \right)^2\,. \label{eqn:Lradio}
\end{align}
Hence, the detected luminosity, e.g., for ILTJ1101+5521, $\mathcal{L}_{\rm radio}^{\rm ILT} = 1.7 \times 10^{27} (d/333\rm pc)^2\, \rm erg \,s^{-1}$ places a strong lower bound on the magnetic moment $\mu$, and the energy conversion efficiency, $\xi$.

The other constraint comes from the fact that any connected fluxtube between the WD and MD is limited by the orbital separation. As a consequence this places a limit by how much the magnetic field can drop inside the flux tube. In turn, radio emission will be dependent on the local value of the magnetic field at the point of emission.

Emission is produced by means of the relativistic electron cyclotron maser instability \citep{1979ApJ...230..621W,1982ApJ...259..844M}, which is driven by a resonance condition between the electrons and the electromagnetic waves, with frequencies $\omega$ and wave vector $k$ such that,
\begin{align}
    \omega - \frac{s}{\gamma_e} \omega_g - k_\parallel v_{e,\parallel} =0\,, 
\end{align}
where $s$ is the harmonic, $\gamma_e$ the electron Lorentz factor, and  $k_\parallel$ and $v_{e,\parallel}$ are the wave vector components and electron velocity along the local magnetic field, respectively. The electron cyclotron frequency is denoted by $\omega_g= e B_e/ \gamma_e m_e c$.  We can see that emission will be primarily produced for $\omega \simeq s \omega_g$ for perpendicular wave vectors (we will further comment on the emission geometry below).

We can estimate the location of the emission as follows. Assuming a source frame emission peak frequency, $\nu_{\rm peak} \simeq s \omega_g/2\pi\gamma_e \simeq \omega_g/2\pi \gamma_e$ for the lowest harmonic $s=1$, which we associated primarily  at the distance $R_e$ of emission from the WD, such that 
\begin{align}
    B_\ast &= B_e \left(\frac{R_e}{R_{\rm WD}}\right)^3 \approx
    \frac{m_e c}{\gamma_e e}\omega_g \left(\frac{R_{e}}{R_{\rm WD}}\right)^3 \,.
    \end{align}
    Since $R_e$ is bounded by the orbital separation,
    \begin{align}
    B_\ast&< \frac{ \gamma_e m_e c}{ e}\omega_g \left(\frac{a + R_{\rm MD}}{R_{\rm WD}}\right)^3 \nonumber\\
    &\approx 2\, \pi\frac{ \gamma_e^2 m_e c}{e}\nu_{\rm peak} \left(\frac{a + R_{\rm MD}}{R_{\rm WD}}\right)^3 \approx 4 \times 10^7 \, \rm G \left(\frac{\gamma_e}{1.4}\right)^2\,, \label{eqn:B_const}
\end{align}
where we have used that $\nu_{\rm peak}\simeq 140\,\rm MHz$. Alternatively, we can express this as $\mu \lesssim 10^{34}\,\rm G\,cm^3$, which is well consistent with the radio luminosity bound.

We get additional constraints from the emission geometry, which in the binary unipolar inductor scenario is intimately tied to the observed time variability seen by a distant observer. In short, since the binary magnetospheric interaction will always be in an {\it on}-state, effective repetition of the signal must be because of short exposure to the emission beam, akin to the pulsation of a radio pulsar. Similar geometrical effects where also recently investigated by~\citet{Barkov:2025uag}.

The emission geometry of the ECMI imposes additional constraints on the system, since 
intrinsically the emission will preferentially occur at a $90^\circ$ angle to the magnetic field line \citep{1969ApJ...156...59G}, as we will also demonstrate in our kinetic simulations.
This has been extensively studied in the context of exoplanet radio emission, since detectability (and observed periodicity) strongly correlates with orbital parameters \citep{2023MNRAS.524.6267K}.

ECMI-powered radiation will be emitted from a point along the magnetic field line under a colatitude $\theta_e$.
However, as seen from the WD, the emission will be beamed to a conical sheet with opening angle,
\begin{align}\label{eqn:aberation}
    \varepsilon = {\rm acos}\, \beta_{e,\parallel}  \approx  \frac{\pi}{2} - {\beta_{e,\parallel}}\,.
\end{align}
It is important to note that the emission will not be uniform across this cone but will be concentrated to a narrow ring-like shape of width $\Delta \varepsilon$ \citep{1985ARA&A..23..169D}.

\begin{figure}
\centering
\includegraphics[width=0.47\textwidth]{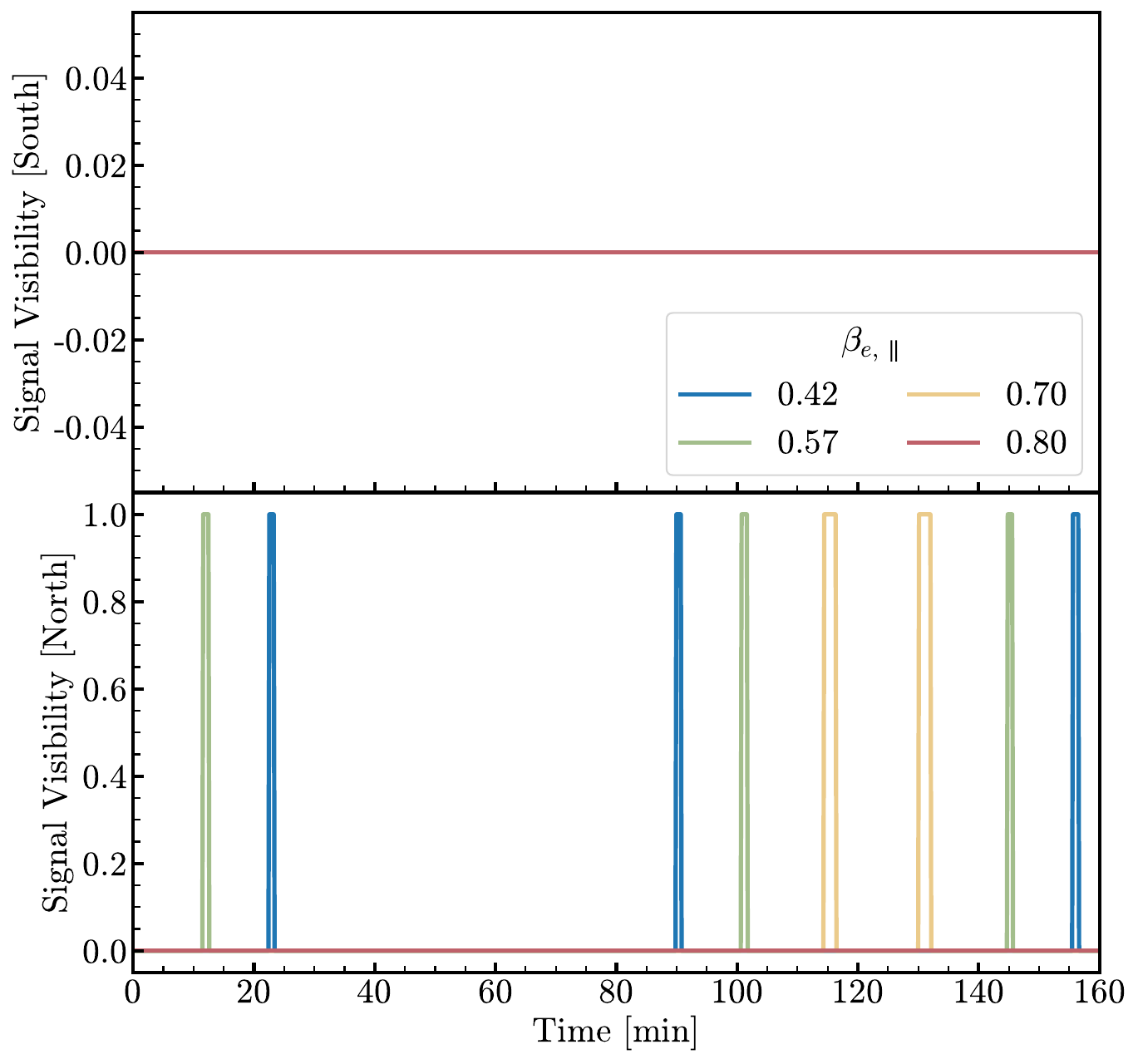}
\caption{Visibility of polarized radio emission as a function of time from the white dwarf -- M-dwarf binary for the parameters considered in this work, generated by \texttt{MASER}~\citep[][]{2023MNRAS.524.6267K}. 
Specifically, we adopt $B_\ast=10^6\,\rm G$, a binary orbital inclination $\iota=20^\circ$, $M_{\rm WD}=0.5\,\rm M_\odot$, $\Delta \varepsilon = 6^\circ$ and a binary period $P=125\,\rm min$.
The top panel shows the visibility of radio emission from the southern hemisphere, and the bottom panel from the northern hemisphere of the white dwarf magnetosphere. The dependence on $\beta_{e,\parallel}=0.42,\,0.57,\,0.70,$ and $0.80$ is represented by the blue, green, yellow, and red solid curves, respectively. We can see that the emission geometry alone naturally leads to minute-long duration pulses with hour-long periods for mildly relativistic velocities.
}
\label{fig:beta_vis}
\end{figure}

We can then use the formalism outlined in \citet{2023MNRAS.524.6267K},
to derive constraints on the bulk velocity, $\beta_{e, \parallel}$, as follows. The orbital inclination, masses, and separation are fixed (or in case of the mass function constrained) observationally.
The magnetic field strength can be bounded using the constraints above, see Eqs. \eqref{eqn:Lradio} and \eqref{eqn:B_const}.
Assuming an aligned dipole, corotation of the WD, and a non-precessing orbit, the period and duration of the radio pulses can be obtained using relativistic aberration \eqref{eqn:aberation}. Specifically, $\Delta \varepsilon$ and $\beta_{e,\parallel}$ will constrain the width of the radio pulse, which can easily be of minute duration and hour-long periods for typical maser parameters.
We demonstrate this in Fig. \ref{fig:beta_vis}, showing that $\gamma_{e,\parallel}\lesssim 1.6$, consistent with a mildly relativistic plasma, which we assume in the following.
We leave a more detailed investigation and parameter inference from the two observed events to future work.

Finally, we need to place a constraint on the electron number density in the radio emitting region. This is important for studying the emission properties, and whether or not the ECMI can operate in the first place.

We do so by estimating the density of current carrying electrons. In the unipolar inductor model, we can estimate the electromotive force, $\mathcal{E}_{\rm EMF} = \sqrt{\dot{E}_{\rm diss} \mathcal{R}/2 }$, where $\mathcal{R}$ is the total resistance, and the factor $2$ arises from the presence of a flux tube in the upper and lower hemisphere \citep{Lai:2012qe}. Assuming that the resistance is largely provided by Spitzer conductivity of the atmosphere plasma \citep{1953PhRv...89..977S},
\begin{align}
    \sigma = \gamma_\sigma \left(\frac{2^{5/2}}{\pi^{3/2}}\right) \frac{\left(k_b T_e\right)^{3/2}}{\sqrt{m_e} Z e^2 \log \Lambda}\,,
\end{align}
here $Z$ is the ion charge number, $T_e$ the electron temperature, $\gamma_\sigma\simeq 0.6$ for $Z=1$, and $\Lambda$ the Coulomb logarithm.
For a WD with $T_e\simeq 10^5\, \rm K$, $\sigma_{\rm WD} \simeq 10^{13}-10^{14}\ \rm s\, cm^{-1}$ (see also \citet{2002MNRAS.331..221W}).
Dissipation now is not just proportional to the resistivity of the object, but also depends strongly on the geometrical properties of the current layer.
\citet{2002MNRAS.331..221W} show that for an arc like dissipative region on the WD (consistent with observations in Jupiter \citep{1993Sci...262.1035C,2024JGRE..12908130M}), and atmospheric dissipation on the secondary, the ratio of the dissipated power between the WD and MD scales as,
\begin{align}
    \frac{\dot{E}_{\rm diss}^{\rm WD}}{\dot{E}_{\rm diss}^{\rm MD}} \simeq \frac{\sigma_{\rm MD}}{\sigma_{\rm WD}} \frac{1}{\delta_{\rm MD}} \left(\frac{a}{R_{\rm WD}}\right)^{3/2} \simeq 3,000 \left(\frac{10^{-3}}{\delta_{\rm MD}}\right)\,, 
\end{align}
where $\delta_{\rm MD} = \Delta R_{\rm MD}/R_{\rm MD}$ is the relative ratio of the atmosphere thickness, $\Delta R_{\rm MD}$, of the MD. We can see that dissipation is predominantly governed by the hotspots on the WD, as the field fall-off suppression dominates over the increase in resistance for the MD.
Following \citet{Willes:2003mw}, we can estimate the resistance of the WD, 
\begin{align}
\mathcal{R}_{\rm WD} = \frac{1}{2 \sigma_{\rm WD}R_{\rm MD}} \left(\frac{a}{R_{\rm WD}}\right)^{3/2} \,.
\end{align}
The resistance will be active in an arc like area, $A=\pi \delta_{\rm arc} R_{\rm tube}^2$ \citep{Willes:2003mw}, where $\delta_{\rm arc}$ is the relative width of the arc region, $R_{\rm tube}\approx R_{\rm MD} (R_{\rm WD}/a)^{3/2}$. We can then relate the electric current inside the circuit, $\mathcal{J} \simeq e v_\parallel A n$, to the dissipated power, yielding a number density near the surface of the WD,
\begin{align}
    n_{\rm WD} &\approx \sqrt{\frac{\dot{E}_{\rm diss}}{2 \mathcal{R}_{\rm WD}}} \frac{1}{\pi e v_\parallel \delta_{\rm arc} R_{\rm tube}^2}\nonumber\,,\\
    &\approx 4 \times 10^{12} {\rm cm}^{-3} \left(\frac{0.1 c }{v_\parallel}\right) \left(\frac{0.01}{\delta_{\rm arc}}\right)\,.
\end{align}

Along the flux tube, the number density of the electrons scales as,
\begin{align}
    n(r) =&\, n_{\rm WD} \left(\frac{R_{\rm WD}}{r}\right)^3\,,\\
    =&\, 3 \times 10^7 {\rm cm^{-3}} \left(\frac{0.8 a}{r}\right)^3\,,
\end{align}
where we have used that the frequency constraint on the emission likely implies an emission zone around $80\%$ of the binary orbital separation.
This results in a magnetization 
\begin{equation}\label{eqn:sigma}
    \sigma \equiv \frac{B^2}{4 \pi n m_e c^2} = \left(\frac{\omega_g}{\omega_{p,e}}\right)^2 \sim 30 \left(\frac{0.8a}{r}\right)^3,
\end{equation}
where $\omega_{p,e}=\sqrt{4\pi e^2 n_e/m_e}$ is the electron plasma frequency. 
This places the emission firmly in a mildly relativistic regime, whereas the surface of the WD will be strongly relativistic.

In summary, we find that the radio emission observed in ULPTs is fully consistent with the orbital motion-driven ECMI in a mildly relativistic regime.

\section{Methods}\label{sec:methods}

{In this work, we perform a detailed investigation of the relativistic ECMI under conditions relevant for WD-MD binary systems outlined in Section \ref{sec:theory}. We do so by numerically solving the special-relativistic Maxwell equations coupled to  special-relativistic collisionless Vlasov system,}
{
\begin{align}\label{eqn:Vlasov}
    p^\mu\partial_\mu f + F^{\alpha \beta} p_\alpha \partial_{p^\beta} f =0\,,
\end{align}
}{
where $f(x^\mu, p^\nu)$ is the particle distribution, $F^{\alpha \beta}$, is the Maxwell tensor, and $x^\mu$ and $p^\mu$ are the particle position and momenta, respectively. In order to initialize the system with an ECMI unstable distribution, we make use of the Dory-Guest-Harris distribution function \citep{1965PhRvL..14..131D,1986ApJ...307..808W}, }
\begin{equation} \label{eq:f}
f(p,\alpha) = {p_\perp}^2(\alpha) \exp\!\left(-\frac{p^2}{(\Delta p)^2}\right),
\end{equation}
{which parametrizes the distribution function, $f$, in terms of a pitch angle $\alpha$, perpendicular momentum $p_\perp = p \sin \alpha$, and a thermal momentum width $\Delta p$, which we obtain from }
\begin{equation} \label{eq:Tth}
    k_\mathrm{B} T_\mathrm{th} = m_e c^2 (1 + (\Delta p)^2)^{1/2} - m_e c^2\,,
\end{equation}
where $T_\mathrm{th}$ is the bulk electron temperature.
Such a distribution is consistent with mirror reflection-driven loss cone distributions \citep{1965PhRvL..14..131D}, and has been used in similar contexts of solar ECMI studies \citep{2021ApJ...909L...5L,2024A&A...681A.113L}.
{In line with several numerical kinetic works on maser emission \citep{Plotnikov:2019zuz,Babul:2020sil,Vanthieghem:2024van}}, we assume primarily an electron-positron composition of the plasma we model (that is $m_i/m_e=1$, with $m_i$ and $m_e$ being the ion and electron masses, respectively), though we comment on the implications of more realistic species mass ratios in Section \ref{sec:polarization}.
{We solve Eq.~\ref{eqn:Vlasov} numerically by using the PIC method \citep{1991ppcs.book.....B} as implemented in the relativistic \texttt{Tristan-MP} PIC code \citep{2005AIPC..801..345S}.}
{We adopt} a two-dimensional Cartesian mesh $(\hat{\boldsymbol{x}}, \hat{\boldsymbol{y}})$ spanning $x, y \in (0, 77)\, c/\omega_p$, where the plasma skin depth is $c/(\Delta x\omega_p) = 20$ with $\Delta x$ being the cell spacing, {but evolve all components of the electric and magnetic fields.} {All boundary conditions in our numerical simulations are imposed to be periodic, though we {\it rejuvenate} particles on the left boundary as outlines below.} {We focus on a local patch along the connected flux tube between the binary companion, where we can neglect spatial curvature of the magnetic field.} Hence, we assume a uniform guiding magnetic field $\boldsymbol{B}_0$, {which we align with the} $\hat{\boldsymbol{x}}$ axis, with a strength chosen such that the plasma magnetization parameter $\sigma_0 \equiv {B_0^2}/({4\pi n_0 m_e c^2}) = 8$, where $n_0 = 16$ denotes the number of macro-particles initialized per cell. 

\begin{table}[ht]
\centering
\setlength{\tabcolsep}{6pt}      
\renewcommand{\arraystretch}{1.1} 
\begin{tabular}{ m{0.31\linewidth} m{0.1\linewidth} m{0.45\linewidth}}
\hline\hline
 & Notation & {~~~ Range} \\ \hline
Temperature [keV] & $T_\mathrm{th}$ &
\{12, 23, 30, 39, 52, 61, 83, 111, 154, 205, 276, 341\} \\
Pitch angle cutoff & $\alpha_\mathrm{c}$ &  $\{0.1, 0.5, 0.9 \}$ \\
\hline
\end{tabular}
\caption{Simulation parameters considered in this work.}
\label{table:1}
\end{table}

The plasma magnetization $\sigma_0$ is chosen such that the Larmor radius of sampled particles is sufficiently large to be numerically resolved ($c/(\Delta x\omega_g)= 7 $), yet small enough for the computational domain to encompass multiple gyrations as the plasma streams with a net drift velocity induced by the pitch-angle cutoff, $\alpha_c$. 
This drift velocity is given by  
\begin{equation} \label{eq:b_shift}
\beta_\mathrm{drift} = \frac{\displaystyle \int_{0}^{p_\mathrm{max}}\!\!\int_{\alpha_c}^{\pi/2}\!\!\int_{0}^{2\pi} f(p,\alpha)\, p\cos\alpha \, d\phi \, \sin\alpha\, d\alpha \, p^2 dp}
{\displaystyle \int_{0}^{p_\mathrm{max}}\!\!\int_{\alpha_c}^{\pi/2}\!\!\int_{0}^{2\pi} f(p,\alpha) \, d\phi \, \sin\alpha\, d\alpha \, p^2 dp}.
\end{equation}

The particle distribution is initialized according to the relativistic loss-cone form given in Eq.~\ref{eq:f}. We further introduce an effective pitch angle cutoff at $\alpha = \alpha_c$. To ensure proper sampling and normalization of the distribution, we employ a rejection sampling scheme as follows:
\begin{enumerate}
    \item Uniformly sample $\cos \alpha$ from the interval corresponding to $\alpha_c \leq \alpha \leq \pi/2$.
    \item Uniformly sample $r \equiv (p/p_\mathrm{max})^5$ for $0 \leq p \leq p_\mathrm{max}$, where the upper bound for the sampled momentum is estimated as $p_\mathrm{max} = 5\,\Delta p$.
    \item Accept the sample $(p,\alpha)$, if it satisfies $u \leq \left( 1 - \cos^2 \alpha \right) \exp\left[ -p^2 / (\Delta p)^2 \right]$, where $u$ is uniformly distributed in $[0, 1]$. Otherwise, reject the sample and repeat the procedure.
\end{enumerate}

For our fiducial case $T_\mathrm{th} = 52$ keV (or equivalently, $\beta_\mathrm{th} \equiv (1-(1+k_\mathrm{B}T_\mathrm{th}/m_e c^2)^{-2})^{0.5}= 0.42$) and $\alpha_c = 0.5$, Eq.~\eqref{eq:b_shift} yields $\beta_\mathrm{drift} = 0.25$, corresponding to an initial bulk streaming velocity of $0.25c$.

\begin{figure*}
    \centering
    \includegraphics[width=1.\linewidth]{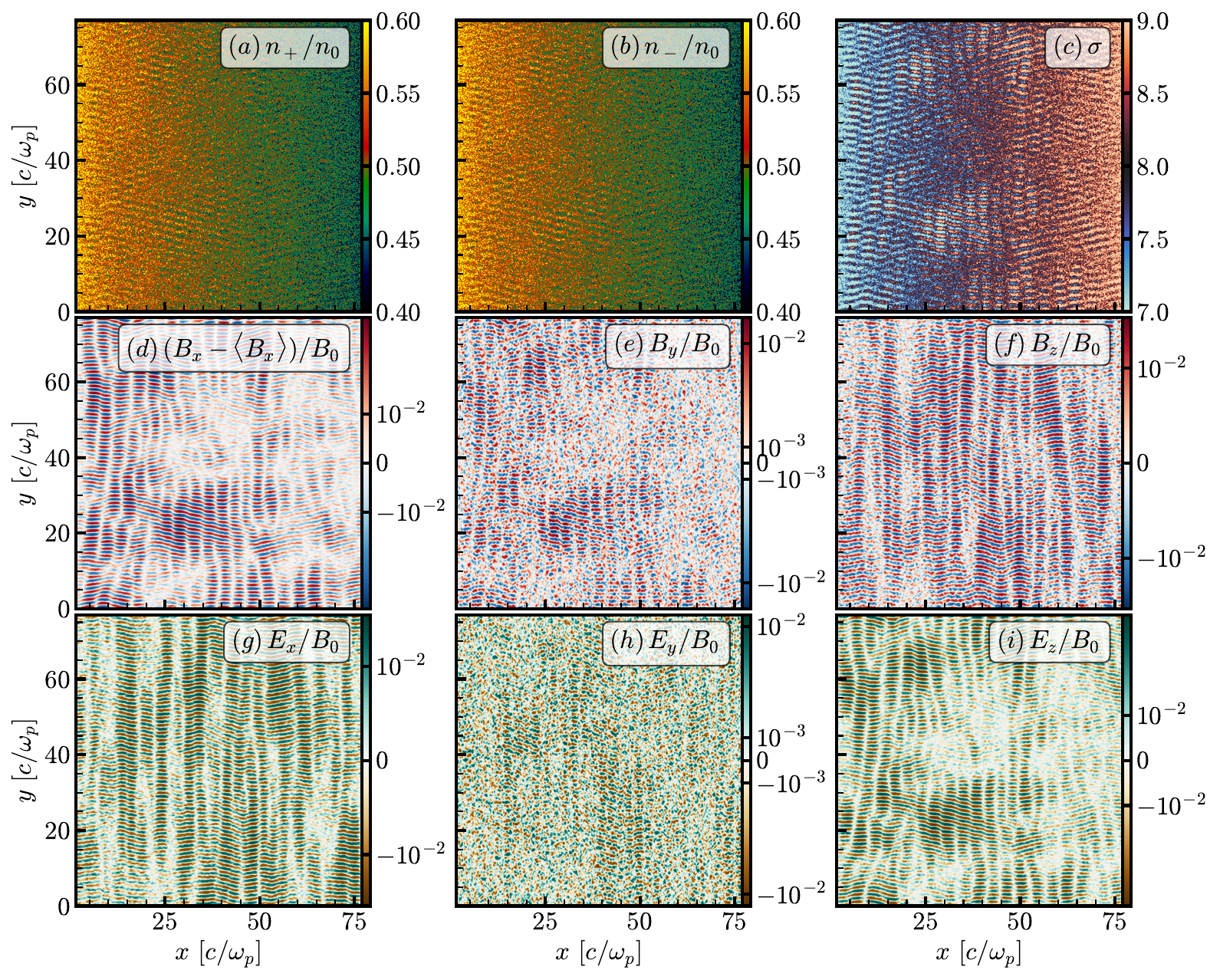}
    \caption{{Steady state of the} electron cyclotron maser instability (ECMI) at the end of the nonlinear growth stage for the fiducial set of parameters $\beta_{\mathrm{th}} = 0.42$ and $\alpha_{\mathrm{c}} = 0.5$. From top left to bottom right, the panels show: (a) positron density normalized to the initial number density, $n_0$, per cell, $n_+/n_0$; (b) normalized electron density, $n_-/n_0$; fluctuation of the parallel magnetic field; (c) plasma magnetization, $\sigma \equiv B^2/4 \pi (n_+ + n_-) c^2$; (d) $(B_x - \langle B_x \rangle)/B_0$, normalized to the initial guiding field $B_0$; (e) normalized in-plane perpendicular magnetic field, $B_y/B_0$; (f) normalized out-of-plane perpendicular magnetic field, $B_z/B_0$; (g) normalized parallel electric field, $E_x/B_0$; (h) normalized in-plane perpendicular electric field, $E_y/B_0$; (i) normalized out-of-plane perpendicular electric field, $E_z/B_0$; Here the spatial coordinates $x$ and $y$ are measured by the plasma skin depth $c/\omega_p$, where $\omega_p$ is the plasma frequency.}
    \label{fig:vth_0.42_alphac_0.5}
\end{figure*}
\begin{figure*}
    \centering
    \includegraphics[width=1.0\linewidth]{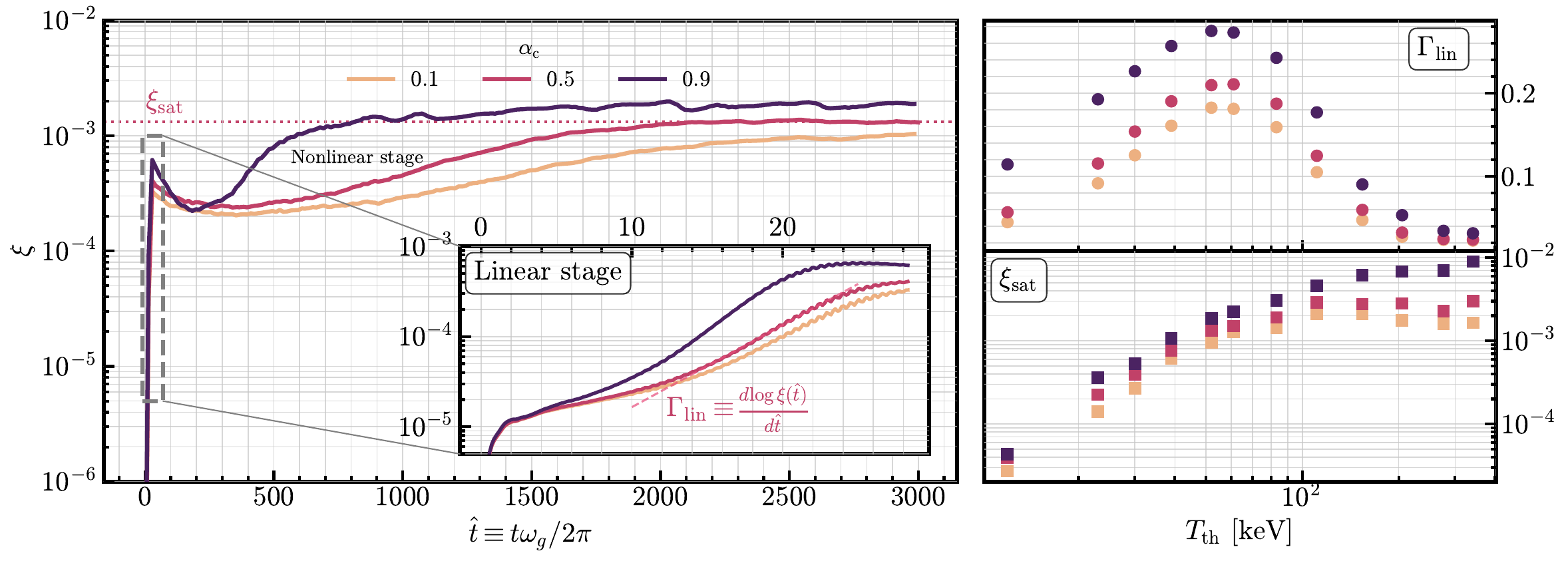}
    \caption{
    Energy conversion efficiency of the relativistic electron cyclotron maser instability, $\xi$, and
    linear growth rate $\Gamma_{\mathrm{lin}}$.
    The saturation value, $\xi_{\mathrm{sat}}$, of the energy conversion efficiency in the nonlinear stead state is also shown. Times, $t$, are given relative to the cyclotron period, $\omega_g/2\pi$, and plasma temperatures, $T_{\rm th}$, are shown in the range of $12\,\mathrm{keV}$ to $341\,\mathrm{keV}$. Different colors denote different pitch angle cutoffs $\alpha_{\mathrm{c}}$. The range of investigated parameters is shown in Table {\ref{table:1}.} }
    \label{fig:gamma_lin_eta_sat}
\end{figure*}

\section{Results} \label{sec:result}

In the following, we discuss our numerical modeling of the ECMI in the regime applicable to the WD-MD binary magnetosphere. 

We begin by describing the emission geometry and properties of the ECMI in the nonlinear saturated quasi-steady regime for a fiducial configuration having $\beta_{\rm th} = 0.42$ and $\alpha_c=0.5$ (Fig. \ref{fig:vth_0.42_alphac_0.5}).
At this stage, the instability starts to saturate and generates a rich set of spatial features in both the electromagnetic fields and the particle densities. 
In the setup we consider (see Section \ref{sec:methods}), the electrons and ions/positrons are {sampled on a uniform} background {magnetic field along} the $x$-direction with field strength $B_0$. We {re-sample the particle momentum near the left boundary of the computational domain according to the Dory–Guest–Harris distribution (Eq.~\ref{eq:f}) to effectively sustain the anisotropy. This continuous injection of anisotropy results in a small gradient in number density}. Correspondingly, the magnetization we model in the steady state regime fluctuates between $7-9$, but not strongly so. 
We can spot coherent planar wavelike patterns in almost all quantities in Fig. \ref{fig:vth_0.42_alphac_0.5} corresponding to the emission of maser radiation. 

The maser instability will be suppressed for alignment angles, $\theta$, of the wave with the magnetic field, where $\left|\cos \theta \right| < \sqrt{k_B T_e/m_e c^2}\ll 1$, in the mildly relativistic regime we consider \citep{1969ApJ...156...59G}. Accordingly, maser radiation will predominantly propagate in the y-direction in our setup, as clearly shown in Fig. \ref{fig:vth_0.42_alphac_0.5}. Consistently, only the $x$- and $z$-components of the electric and magnetic field exhibit a clear propagation pattern of the waves. In terms of plasma waves, these two different components correspond to the $O$- and $X$-mode, respectively. 
The $O$-mode consists of electric field components parallel to the background magnetic field, so that we may write $\xi_O = \left<B_z^2\right>/ \left<B_0^2\right>$ for its relative energy, and similarly $\xi_X = \left<\left(B_x - B_0 \right)^2\right>/ \left<B_0^2\right>$ for the $X$-mode. The $X-$mode can propagate approximately above the electron cyclotron frequency, $\omega_X = (\omega_g + \sqrt{\omega_g^2 + \omega_{p,e}^2})/2 = \omega_g \left(1+ \sqrt{1 + 1/\sigma_0}\right)/2\approx \omega_g$, whereas the $O-$mode propagates above the plasma frequency, $\omega_{p,e}$ \citep{1982ApJ...259..844M}.

We {observed} that $X-$mode growth is the most rapid, as expected from linear theory \citep{1982ApJ...259..844M}, as well as seen in related simulations of radiatively cooled laser plasmas \citep{2025SciA...11.8912B}.
In the saturated state, we roughly find that the ratio of O to X-Mode energies for our fiducial case is about $\xi_O/\xi_X \simeq {0.2}$. The different growth of these modes will also have strong implications for the polarization content of the resulting emission, as we will discuss below. 
It is interesting to ask whether any of these waves could potentially alter the magnetic field background geometry itself \citep{2006ApJ...652.1297L,2017ApJ...840...52I,Sobacchi:2024iyw,Sobacchi:2024yis,Beloborodov:2025oei}. 
This can be computed by considering the wave strength parameter,
\begin{align}
a = \frac{e\, \delta E_z}{m \omega}
\approx \frac{\omega_g}{\omega}
\frac{\delta E_z}{B_x} \lesssim \frac{\delta E_z}{B_x} \simeq 10^{-2}\ll 1  \,,
\end{align}
where the first inequality follows from the dispersion relation of the ECMI, and the second inequality from our results. As a result, unlike in the shock-powered synchroton maser context \citep{Plotnikov:2019zuz,Sironi:2021wca}, the ECMI in the WD-MD regime we consider never feeds back on the background magnetic field.

We can further see that there is another longitudinal mode seemingly propagating in the $x$-direction with larger wavelength. This is consistent with the presence of a trapped $Z-$mode~\citep{1986ApJ...307..808W}. 
In the following, we do not filter out the Z-mode from our analysis, though we caution that nonlinear decay of this mode, e.g., into an X-mode, is necessary for it to produce observable radiation \citep{2021ApJ...909L...5L}.

One of the most important questions of this work, which can only be carried out by means of numerical simulations, is the efficiency of the maser operating in the common WD-MD magnetosphere. Since the observed radio luminosity places strong limits on the emission mechanism, it is important to understand how efficiently the ECMI can convert the available energy into EM radiation. As such, we use a commonly employed measure for energy conversion efficiency \citep{Plotnikov:2019zuz,Babul:2020sil}, 
\begin{equation}
    {\xi} = \frac{\int{\rm d}V\left( \delta B^2 + \delta E^2\right)}{\int {\rm d}V B_0^2} \simeq \xi_O + \xi_X\,,
\end{equation}
which relates the wave energy density, $e_{\rm wave} = \left( \delta B^2 + \delta E^2\right)/2$, to the background magnetic field energy density, $e_{\rm bg}= B_0^2/2$.

While we do not differentiate between trapped an non-trapped modes in this approach, it provides a good and convenient estimate of how efficiently the maser can operate given a WD magnetic field.
We then evaluate $\xi$ for different combinations of $(\beta_\mathrm{th},\alpha_\mathrm{c})$. As a representation, we show the time evolution curves $\xi(\hat{t})$ ($\hat{t} \equiv t \omega_g / 2 \pi$ is the time normalized by the gyration timescale defined by plasma cyclotron frequency $\omega_g$) for three pitch angle cutoffs under the fiducial thermal velocity we considered ($\beta_\mathrm{th}=0.42$) in the left panel of Fig.~\ref{fig:gamma_lin_eta_sat}, where the purple, magenta and orange curves stand for $\alpha_\mathrm{c}=0.9,0.5, 0.1$, respectively. Here one can clearly see two growing stages: the linear one at $\hat{t} \lesssim 50$ and the nonlinear one at $\hat{t} \gtrsim 300$. To characterize the time evolution of the ECMI, we measure the saturated energy efficiency at the end of the nonlinear growing stage $\xi_\mathrm{sat}$ (e.g., for $\alpha_c=0.5$, it is denoted by the dotted line in the left panel), as well as the linear growth rate,
\begin{equation}
    \Gamma_\mathrm{lin} \equiv \frac{d \log \xi(\hat{t})}{d\hat{t}} \approx \frac{d \log \xi_X(\hat{t})}{d\hat{t}} ,
\end{equation}
as shown in the inset of the left panel. 
It is important to note that the maximum linear growth rate will roughly happen for $\omega_{\rm max}^s = s \omega_g \left(1 + \beta_{\rm th}^2 \cos^2 \alpha_c\right)$ \citep{1982ApJ...259..844M}, where $s$ is the maser harmonic. Since in the regime we consider, $\omega_{\rm max}^s/\omega_x \gtrsim 1$, the onset of instability for the fundamental $X-$mode is always allowed, and will dominate the growth rate.

The dependence of $\Gamma_\mathrm{lin}$ and $\xi_\mathrm{sat}$ on plasma thermal velocity, or equivalently, the thermal temperature of plasma $T_\mathrm{th}$ \eqref{eq:Tth} are depicted in the top right and bottom right panels, respectively. Here the data points in different colors correspond to different values of the pitch-angle parameter $\alpha_{\mathrm{c}}$. One can see that the ECMI optimally grows in the intermediate temperature domain $30~\mathrm{keV} \lesssim T_\mathrm{th} \lesssim 100$ keV, agreeing with~\citep{2006A&ARv..13..229T}. The saturated energy efficiency in this regime, on the other hand, roughly follows a power-law trend. 

Most importantly, we find that linear growth of the $X-$mode is followed by a secondary nonlinear growth, in which also the $O-$mode fully develops, though in the regime we explore, $\xi_O \ll \xi_X$, always.
The nonlinear growth rate depends strongly on the pitch angle anisotropy, and can take up to 100-times longer to saturate compared to the linear growth (Fig. \ref{fig:gamma_lin_eta_sat}).
We confirmed with a separate set of simulations not presented here, which did not include continued anisotropy injection (i.e., a decaying maser set up), that the nonlinear growth is absent and the instability decays as the plasma thermalizes.

Overall, we find that the maser efficiency can be as high as $\xi\simeq 10^{-2}$, with the nonlinear stage contributing up to an order of magnitude correction to this value. \\

\begin{figure}
\centering
\includegraphics[width=0.47\textwidth]{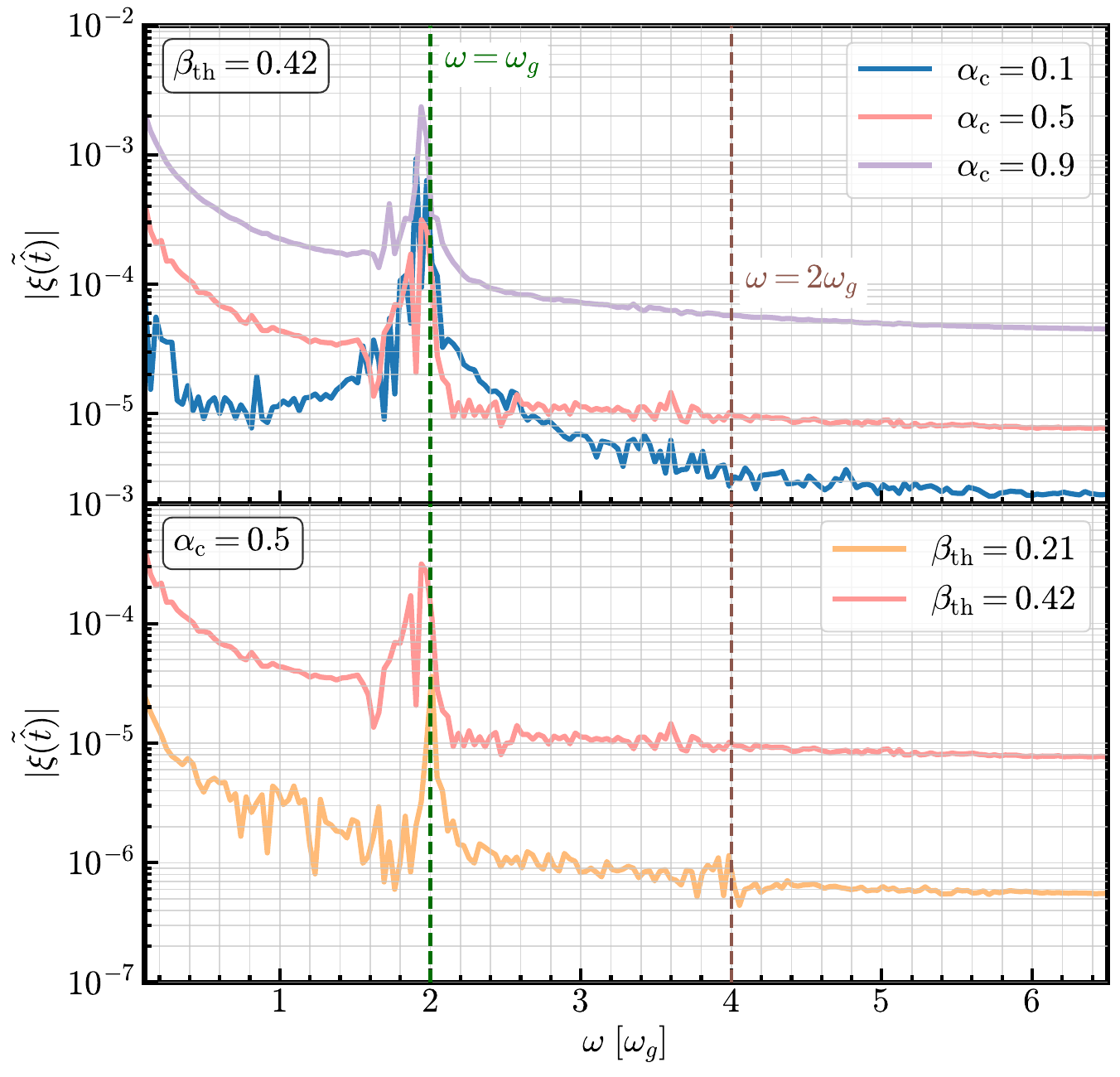}
\caption{Frequency spectrum of the energy efficiency $\xi(t)$. \textbf{Top Panel:} The plasma thermal velocity is fixed at the fiducial value $\beta_\mathrm{th} = 0.42$. To examine the dependence of the emission bandwidth $\Delta \omega$ on the velocity pitch angle, we compare the fiducial case with $\alpha_\mathrm{c} = 0.5$ (red solid curve) to two additional cases with smaller and larger pitch angles: $\alpha_\mathrm{c} = 0.1$ (blue solid curve) and $\alpha_\mathrm{c} = 0.9$ (purple solid curve), respectively.
\textbf{Bottom Panel:} The pitch angle cutoff is fixed at $\alpha_\mathrm{c} = 0.5$, and we compare two cases with different plasma thermal velocities: $\beta_\mathrm{th} = 0.42$ (red solid curve; corresponding to $k_\mathrm{B} T_\mathrm{th} = 50$~keV) and $\beta_\mathrm{th} = 0.21$ (orange solid curve; corresponding to $k_\mathrm{B} T_\mathrm{th} = 10$~keV). The expected resonance at $\omega = \omega_g$ and $\omega = 2\omega_g$ are indicated by the green and brown dashed lines, respectively. {Since $\xi$ scales with the square of the electromagnetic field, a component at $\omega_g$ in the field spectrum appears at $\omega = 2\omega_g$ in $\xi$.}}
\label{fig:eta_t_fourier}
\end{figure}

Observationally, it is important to understand the intrinsic spectral width of the maser emission. We therefore ask what is the energy spectrum per frequency band. 
To extract the energy spectrum of ECMI radiation, we perform a Fourier transform of $\xi(\hat{t})$, with its dependence on $(\beta_\mathrm{th}, \alpha_\mathrm{c})$ shown in Fig.~\ref{fig:eta_t_fourier}. The spectral intensity $|\tilde{\xi}(t)|$ is primarily governed by the plasma thermal velocity, $\beta_{\rm th}$, rather than the pitch-angle cutoff, in agreement with the right two panels of Fig.~\ref{fig:gamma_lin_eta_sat}. In all cases, the spectra peak at $\omega = \omega_g$, indicating that the emission power is concentrated at the fundamental cyclotron resonance, as expected. Higher-order resonance peaks appear only for the case with a low plasma thermal velocity, $\beta_\mathrm{th} = 0.21$ ($T_\mathrm{th} = 12$~keV), where both the linear growth rate and the saturated energy efficiency are suboptimal. This is to be expected in the regime $\omega_g > \omega_{p,e}$ that we probe, since the growth of the fundamental $X-$mode always saturates fastest \citep{1984JGR....89..897M}.

For our fiducial case—coinciding with the maximum linear growth rate as a function of plasma thermal velocity—the emission bandwidth is approximately $\Delta \omega \sim 0.2\,\omega_g$, which is rather narrow compared to, e.g., shock-powered synchrotron maser cases \citep{Sironi:2021wca,Babul:2020sil,Plotnikov:2019zuz}.
\begin{figure*}
    \centering
    \includegraphics[width=1.0\linewidth]{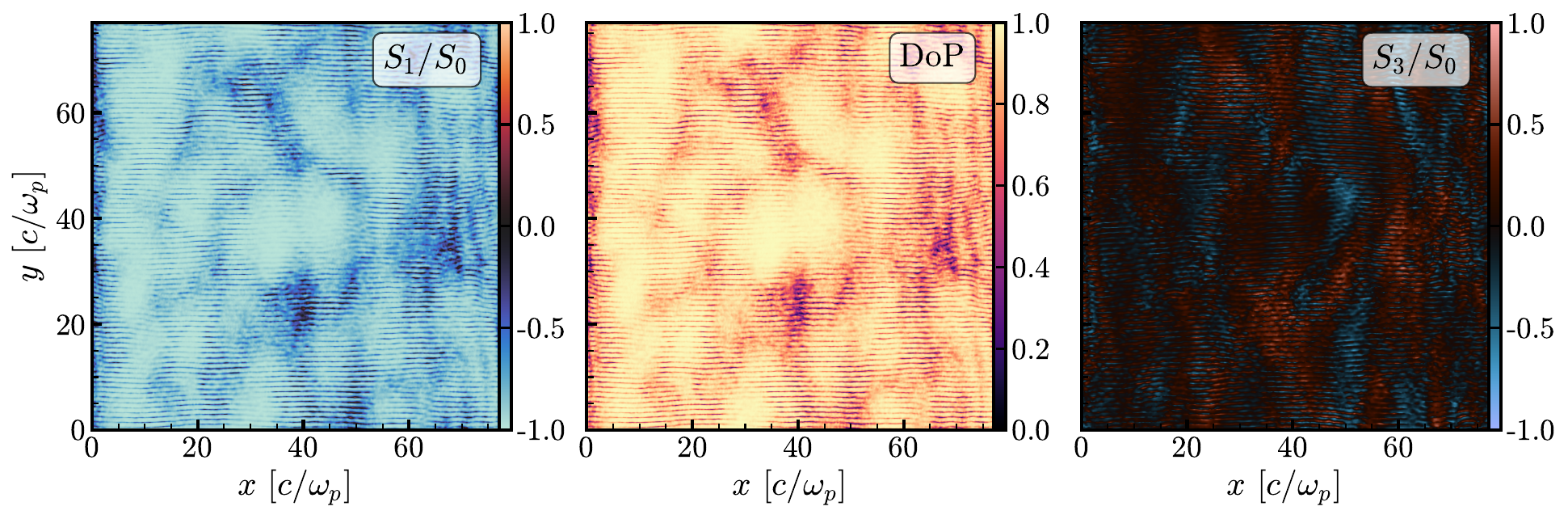}
    \caption{Polarization properties of the ECMI for a plasma thermal velocity $\beta_{\mathrm{th}} = 0.42$ (corresponding to $k_{\mathrm{B}} T_{\mathrm{th}} = 52~\mathrm{keV}$) and a velocity pitch angle $\alpha_{\mathrm{c}} = 0.5$. From the top left to the bottom right panels, we show the three Stokes parameters normalized by the total intensity, the total degree of polarization, the phase difference between $E_{\mathrm{y}}$ and $E_{\mathrm{z}}$, and the ratio of linear to circular polarization.}
    \label{fig:polarization}
\end{figure*}

\subsection{Polarization}\label{sec:polarization}

Utilizing the two transverse electric field components of our fiducial case, $E_y(t)$ and $E_z(t)$, at the end of nonlinear growth stage, we characterize the polarization state of the emitted radiation using the Stokes parameters $S_i$, together with the degree of polarization (DoP) 

To access both amplitude and phase information, we construct the complex analytic signals, $\tilde{E}_y(t)$ and $\tilde{E}_z(t)$, by applying the Hilbert transform to the real-valued simulation outputs. 
In particular, we focus on the total time-averaged intensity
\begin{equation}
    S_0 = \left\langle |\tilde{E}_y|^2 + |\tilde{E}_z|^2 \right\rangle ,
\end{equation}
the linear polarization along the simulation $y$- and $z$-axes
\begin{equation}
    S_1 = \left\langle |\tilde{E}_y|^2 - |\tilde{E}_z|^2 \right\rangle ,
\end{equation}
and the circular polarization
\begin{equation}
    S_3 = 2 \left\langle \mathrm{Im} \left[ \tilde{E}_y^* \tilde{E}_z \right] \right\rangle .
\end{equation}
From these, we can define the degree of linear polarization,
\begin{align}\label{eqn:Pi_linear}
    \Pi  = \frac{S_1}{S_0} \simeq \frac{\xi_X - \xi_O}{\xi_X+\xi_O} \approx 1- 2 \frac{\xi_O}{\xi_X} + \mathcal{O}\left(\left[\frac{\xi_O}{\xi_X}\right]^2\right)\,,
\end{align}
where in the second equality we have used the properties of the Hilbert and Fourier transforms, and in the final approximation that in the regime we consider $\xi_O \ll \xi_X$.

In addition, we define the degree of circular polarization, 
\begin{align}
    \mathcal{C} = \frac{S_3}{S_0}\,
\end{align}
and the total degree of polarization,
\begin{equation}
    \mathrm{DoP} = \frac{\sqrt{S_1^2 + S_2^2 + S_3^2}}{S_0}\,,
\end{equation}
where
\begin{equation}
    S_2 = 2 \left\langle \mathrm{Re} \left[ \tilde{E}_y^* \tilde{E}_z \right] \right\rangle
\end{equation}
is the linear polarization at $\pm 45^\circ$. The resulting two-dimensional maps of the polarization diagnostics in Fig.~\ref{fig:polarization}.

In the WD-MD regime we consider, the growth of the fundamental $X-$mode always dominates the $O-$mode growth. As a result, $\xi_O/\xi_X \ll 1$, leading to strong linear polarization fractions of $\Pi \simeq 0.7-0.8$, dominating the DoP (Fig. \ref{fig:polarization}). 
This is in stark contrast to the regimes probed in planetary radio emission, which is predominantly circularly polarized \citep{2006A&ARv..13..229T}. Here we only find $\mathcal{C}\simeq 0.2$.
These values are consistent with the polarization fractions observed in ILT J1101$+$5521 ($\Pi_{\rm Obs} \simeq 0.2-0.5$, $\mathcal{C}_{\rm Obs} \lesssim 0.05$; \citet{deRuiter:2024yib}) and GLEAM$-$X J0704-37 ($\Pi_{\rm Obs} \simeq 0.2-0.5$, $\mathcal{C}_{\rm Obs} \simeq 0.1-0.3$; \citet{2024ApJ...976L..21H}).
 
However, we caution that the details of the polarization will change with varying electron-ion mass ratio. Here, we have considered primarily pair plasmas ($m_e=m_i$, where $m_i$ is the ion mass), in which the circular polarization components can more efficiently cancel due to the symmetry between positrons and electrons when initiated with identical distributions. Turning to the related cases of shock-powered synchrotron maser emission, it has been shown that for sufficient ion $\sigma$ and $m_i/m_e = 200$, the efficiency and linear polarization properties of the emitted waves in a pair plasma \citep{Sironi:2021wca} continue to hold \citep{Iwamoto:2023fxo}. Follow-up work will be required to demonstrate whether similar conclusions hold for the ECMI, as well.

\section{Conclusion}\label{sec:conclusions}

In this work, we have investigated the viability of the ECMI to explain radio emission from LPTs for interacting WD-MD binaries. The observationally driven need to explain the expected high energy conversion efficiency, as well as substantially linear polarization signatures \citep{deRuiter:2024yib,2024ApJ...976L..21H} required a full numerical exploration of the ECMI in the driven nonlinear regime, not previously considered in the literature.

Using kinetic plasma simulations, we demonstrated that the ECMI can efficiently operate under realistic WD-MD conditions, which we estimated using a unipolar inductor scenario \citep{1969ApJ...156...59G,Lai:2012qe} applied to WD binaries \citep{2002MNRAS.331..221W,Willes:2003mw}.
We show that under these mildly relativistic conditions, the ECMI can reach high energy conversion efficiencies, $\xi \simeq 10^{-3}-10^{-2}$, relative to the background magnetic energy density. 
We find that the polarization (in the pair plasma approximation we consider) can be substantially linear ($\Pi \simeq 0.6$), but we caution that this may depend electron-ion mass ratio in the system, with work in the shock-powered synchrotron maser context demonstrating that (under high enough ion magnetization) linear polarization can be maintained \citep{Iwamoto:2023fxo}.\\

Overall, our work demonstrates that the ECMI can efficiently operate in the WD-MD context and can explain the observed radio emission properties. In order to be fully predictive especially in terms of the radio polarization, several steps are necessary. First, similar investigations in the context of shock-powered synchrotron maser emission have shown that the growth of the O-mode (which we find to be primarily a nonlinear effect) may be altered in three-dimensional simulations \citep{Sironi:2021wca}. Second, while linear polarization is easy to produce in the pair context, it the impact of realistic electron-ion mass ratio will be important to quantify. 

Finally, some LPT systems will not necessarily be powered by orbital motion of the binary, but also by wind interaction \citep{Horvath:2025sze}. It may be interesting to consider applying similar kinetic techniques to these systems as well \citep{Zhong:2024gnv}.

\begin{acknowledgments}

The authors are grateful for technical assistance and insightful discussion to Anatoly Spitkovsky and Muni Zhou, and for insightful discussions to Dongzi Li, Alexander Philippov, E. Sterl Phinney, Lorenzo Sironi, and Christopher Thompson. YZ acknowledges support as a Sherman Fairchild postdoctoral prize fellow at the Walter Burke Institute for Theoretical Physics. ERM acknowledges partial support from the National Science Foundation through AST-2307394.
This research was supported in part by grant NSF PHY-2309135 to the Kavli Institute for Theoretical Physics (KITP).

The authors are pleased to acknowledge that the work reported on in this paper was substantially performed using the Princeton Research Computing resources at Princeton University which is consortium of groups led by the Princeton Institute for Computational Science and Engineering (PICSciE) and Office of Information Technology's Research Computing.
ERM further acknowledges support through
DOE NERSC supercomputer Perlmutter under grants m4575 and m5801, which uses resources of the National Energy Research Scientific Computing Center, a DOE Office of Science User Facility supported by the Office of Science of the U.S. Department of Energy under Contract No. DE-AC02-05CH11231 using NERSC award NP-ERCAP0028480. ERM acknowledged support on the NSF Frontera supercomputer under grant AST21006, and on Delta at the National Center for Supercomputing Applications (NCSA) through allocation PHY210074 from the Advanced Cyberinfrastructure Coordination Ecosystem: Services \& Support (ACCESS) program, which is supported by National Science Foundation grants \#2138259, \#2138286, \#2138307, \#2137603, and \#2138296.

\end{acknowledgments}

\software{
    \texttt{Tristan-MP} \citep{2005AIPC..801..345S},
    matplotlib \citep{matplotlib},
    numpy \citep{numpy},
    scipy \citep{scipy},
    \texttt{MASER} \citep{2023MNRAS.524.6267K}
}


\bibliography{sample701}{}
\bibliographystyle{aasjournalv7}

\end{document}